\documentclass[11pt,preprint]{aastex}
\usepackage[scaled]{helvet}
 
\usepackage[T1]{fontenc}
 \usepackage[bookmarks=false]{hyperref}
\usepackage{amsmath}
\usepackage{url}
\begin{document}
\title{Pareidolic Dark Matter (PaDaM)}

\author{ \small N. Mirabal\\
Dpto. de F\'isica At\'omica, Molecular y Nuclear\\
Universidad Complutense de Madrid, Spain\\
}

\bigskip


{\setlength{\parindent}{0cm}
Dark matter is a sphinx \citep{oort, zwicky,rubin,tyutchev}.
Despite ingenious hunting strategies, we continue to 
chase after this notoriously elusive creature within the dark zoo 
\citep{feng,fan}. 
In the  sky, 
there are X-ray ``mice'' \citep{gaensler}, radio ``snakes'' \citep{uchida}, 
even an optical  ``Crab''
\citep{rosse,mitchell}. 
The latter may have been a ``pineapple'', as a matter of
fact \citep{dewhirst}. 
But such discussion is beyond the scope of our tract. }

\noindent
The Universe hosts
astronomical animals aplenty \citep{dacke}. 
However, a dark matter creature should be unique in its peculiarity.  
When finally corralled, 
we conjecture pareidolically 
that -- under extremely finely tuned conditions -- 
dark matter sub-substructures
might resemble   
a ``winged horned lion'' with a  
``serpent'' for a tail,
whose contours become apparent
roughly 
at $q$ times the free streaming scale  
$\lambda_{fs}$ (where $q$ is real or imaginary \citep{green}).
The color figure shows a generic example of 
pareidolia\footnote{\tiny http://dictionary.reference.com/browse/pareidolia}
right 
at the intersection of annihilation and 
diffuse emission \citep{cuesta}. 
The image corresponds to a
simulated {\it Fermi} LAT gamma-ray map obtained
after exhaustive visualization. 
Observers could contemplate 
verifying the truthiness or falsity of this proposition (or not).

\begin{center}
\includegraphics[width=0.58\linewidth]{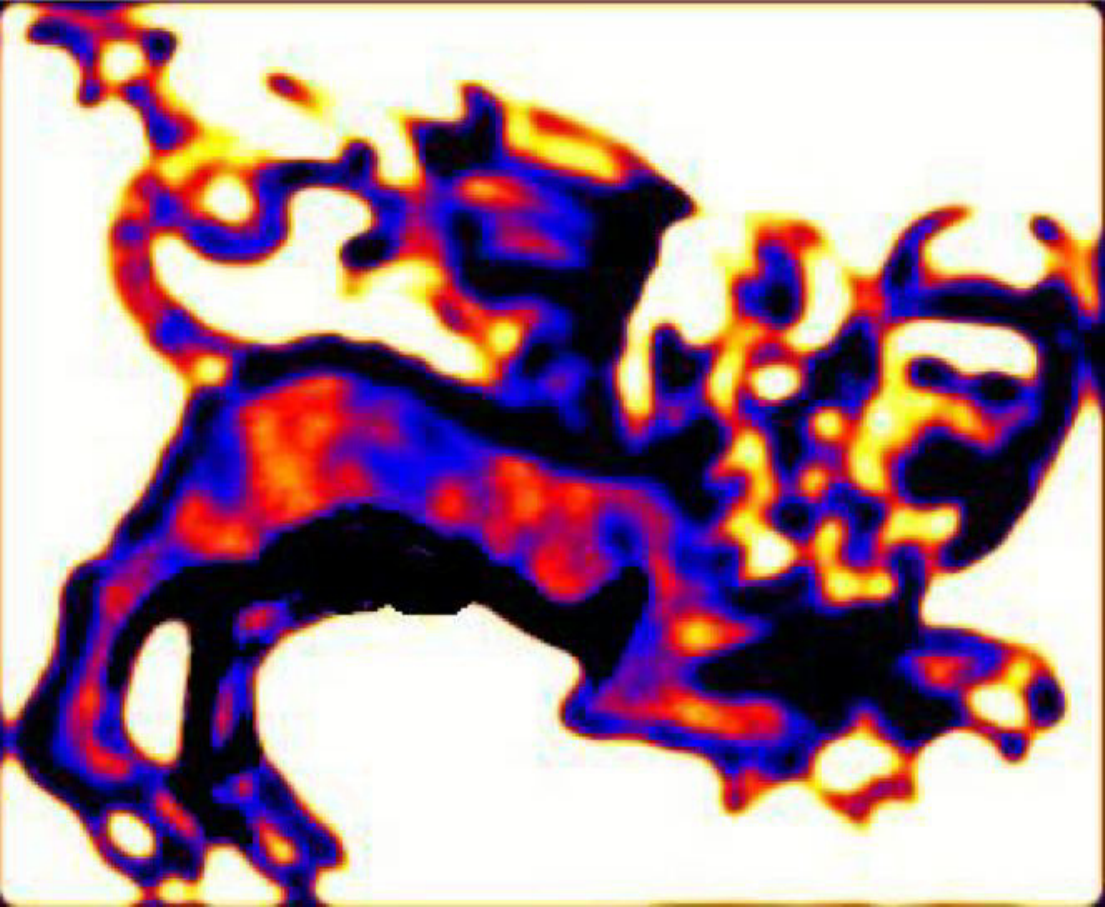}
\end{center}

\noindent
Incidentally,
it seems stark, staringly 
clear that the dark matter particle cannot linger in obscurity for much longer.
Now for those who have followed us so far, it should be 
deceivingly obvious that this is an April Fools' Day spoof. 
Enjoy 
and dear reader, please forgive us if we have wasted your time.

\bigskip

\acknowledgements 
We acknowledge Mark Hurn for kindly sending us an original offprint
of ``Early drawings of Messier 1: pineapple or
crab?'' by the late D.W. Dewhirst. For more astronomical pareidolia  
please visit the Hubble Heritage Creative Challenge at 
\url{https://www.facebook.com/media/set/?set=a.10151117646584962.451794.365958474961&type=1&l=069cb8ba1e}.


\bigskip
\bigskip
\bigskip

\begin{thebibliography}{}
\bibitem[Cuesta et al.(2011)]{cuesta}Cuesta A. J. et al. 2011, 
ApJ, 726, L6

\bibitem[Dacke et al.(2013)]{dacke}Dacke M. et al. 2013, Current Biology,
\url{http://dx.doi.org/10.1016/j.cub.2012.12.034}

\bibitem[Dewhirst(1983)]{dewhirst} Dewhirst D. W. 1983, The Observatory,
103, 114

\bibitem[Fan et al.(2013)]{fan}Fan J., Katz A., Randall L., \& Reece M. 2013,
arXiv:1303.3271

\bibitem[Feng(2010)]{feng}Feng J. L. 2010, ARA\&A, 48, 495

\bibitem[Gaensler et al.(2004)]{gaensler}Gaensler B. M. et al. 2004, ApJ, 
616, 383

\bibitem[Green et al.(2005)]{green}Green A. M. et al. 2005, JCAP, 
08, 003

\bibitem[Mitchell(1855)]{mitchell}Mitchell R. J. 1855,
Scientific Transactions of the Royal Dublin Society, New Series Vol. II. 
plate II


\bibitem[Oort(1932)]{oort}Oort, J. H. 1932, Bull. Astr. Inst. 
Netherlands, 6, 249

\bibitem[Rosse(1844)]{rosse}Rosse, Earl of 1844, Phil. Trans.,
Part II, read on June 13


\bibitem[Rubin \& Ford(1970)]{rubin}Rubin V. C., \& Ford W. K. 1970, ApJ,
159, 379

\bibitem[Tyutchev(1869)]{tyutchev}Tyutchev F. I. 1869, Nature is a Sphinx, From
the Ends to the Beginning: A Bilingual Anthology of Russian Poetry
\url{http://max.mmlc.northwestern.edu/~mdenner/Demo/index.html}

\bibitem[Uchida et al.(1992)]{uchida}Uchida K. et al. 1992, ApJ, 104, 
1533

\bibitem[Zwicky(1933)]{zwicky}Zwicky, F. 1933, Helvetica Physica Acta 6, 110
\end{thebibliography}
\end{document}